\documentclass[oldversion]{aa}
\usepackage{graphicx}
\def\ltsima{$\; \buildrel < \over \sim \;$}
\def\simlt{\lower.5ex\hbox{\ltsima}}
\def\gtsima{$\; \buildrel > \over \sim \;$}
\def\simgt{\lower.5ex\hbox{\gtsima}}
\def\cgs{{erg cm$^{-2}$ s$^{-1}$}}
\def\ergs{{erg s$^{-1}$}}
\def\cm2{{cm$^{-2}$}}
\def\xdof{{$\chi^{2}$(dof)}}
\def\xndof{{$\chi^{2}_{\rm \nu}$(dof)}}
\def\xnu{{$\chi^{2}(\nu)$}}

\def\fhx{{$F_{2-10}$}}

\def\lums{{$L_{0.5-2}$}}
\def\p1{{Paper I}}

\def\xmm{{\em XMM--Newton}}
\def\chandra{{\em Chandra}}
\def\beppo{{\em BeppoSAX}}

\def\asca{{\em ASCA}}

\def\nh{{N$_{\rm H}$}}

\def\nhp{{N$_{\rm H}^{\rm p}$}}
\def\chandra{{\em Chandra}}

\def\xmm{{\em XMM--Newton}}

\def\nh{{N$_{\rm H}$}}

\def\epic{{\em EPIC}}

\def\pn{{PN}}
\def\mos{{MOS}}
\def\rgs{{RGS}}

\def\due{{10$^{22}$}}
\def\tre{{10$^{23}$}}
\def\f14{{10$^{-14}$}}
\def\f13{{10$^{-13}$}}

\def\ngc{{NGC~7582}}
\def\refltot{{{\it refl}}}

\def\feka{{Fe K$\alpha$}}

\begin{document}

\title{XMM-Newton broad-band observations of NGC~7582:\\ N$_{\rm H}$
  variations and fading out of the active nucleus}

\author{E.~Piconcelli\inst{1}, S.~Bianchi\inst{2,3}, M.~Guainazzi\inst{2},
  F.~Fiore\inst{1}, M.~Chiaberge\inst{4}}

\titlerunning{XMM-Newton observations of NGC~7582}\authorrunning{E.~Piconcelli et al.}

\offprints{piconcelli@oa-roma.inaf.it}

\institute{Osservatorio Astronomico di Roma (INAF), via Frascati 33, I--00040
  Monteporzio Catone, Italy \and European Space
  Astronomy Center of ESA, Apartado 50727, E--28080 Madrid,
  Spain \and Dipartimento di Fisica, Universit\`a degli Studi Roma Tre, via
  della Vasca Navale 84, I--00146 Roma, Italy
\and Space Telescope Science Institute, 3700 San Martin Drive, MD
  21218, USA}

   \date{}


\abstract{We present results from two \xmm~observations of the  bright
classical Seyfert 2 galaxy NGC~7582 taken four years apart (2001 May and 2005
April).  We present the analysis of the
high-resolution (0.3-1 keV) \rgs~and low-resolution (0.3-10 keV)
\epic~spectroscopic data.
A comparison with a 1998 \beppo~observation suggests that \xmm~caught the
source in a ``reflection-dominated'' phase, measuring the lowest continuum
flux level ever (\fhx~= 2.3 $\times$ 10$^{-12}$ \cgs) in 2005. \ngc~therefore
experienced a dramatic spectral transition most likely due to the partial
switching-off of the nuclear activity.  The \xmm~spectrum of the continuum
emission is very complex.  It can be well described by  a  model consisting of
a combination of a heavily absorbed (\nh~$\sim$ 10$^{24}$ \cm2) power law and
a pure reflection component both obscured by a column density of $\sim$ 4
$\times$ 10$^{22}$ \cm2.  Notably, we detect  a significant increase by a
factor of $\sim$ 2 in the column density of the inner, thicker absorber covering
the primary X-ray source between 2001 and 2005.  The 2005 \xmm~spectra show the
strongest \feka~emission line 
ever measured in this source. This is consistent with the line delayed
time response to the decrease of the nuclear activity.
Our analysis also reveals that the soft
X-ray spectrum is dominated by emission lines from  highly ionized
metals. The detection of a narrow OVIII radiative recombination continuum
suggests an origin in a photoionized plasma.

}

   \maketitle
%

\section{Introduction}

 In Type 2 Seyfert galaxies the otherwise overwhelming primary continuum
 emission from the active nucleus is attenuated by the presence of an
 obscuring screen (the putative dusty torus of the unified model,
 e.g. Antonucci 1993). This favorable condition enables us to investigate in
 detail the geometrical and physical properties of the emitting/absorbing gas
 in the nuclear environment. X-ray
 observations of obscured sources permit to efficiently map the inner
 kpc-scale region around the accreting black hole (BH) and understand the
 physical processes at work in AGN. Recent \xmm~and \chandra~results have
 provided unprecedented insights  into the structure and dynamics of the
 reprocessing materials (see e.g. Risaliti et al. 2005; Iwasawa et al. 2003;
 Guainazzi \& Bianchi 2007 and references therein) and the starburst
 contribution to the total X-ray emission (e.g. Jimenez-Bail\'on et al. 2003).\\

In this paper we present the analysis of two \xmm~observations of \ngc. This
source is a highly inclined ($i$ $\sim$ 65$^\circ$) barred spiral galaxy at
$z$ = 0.0053 (1 arcsec $\sim$ 110 pc in the source rest-frame).
Optical observations reveal  a complex morphology, with a
kpc-scale disk of H II regions and a prominent dust lane, parallel to the
major axis of the galaxy, running across the nuclear region at a distance of
some hundreds parsecs (Regan \&  Mulchaey 1999; Prieto et al. 2002; Wold \&
Galliano 2006).  The optical spectrum is a superposition of a Seyfert 2
nucleus and a starburst (Cid Fernandes et al. 1998, Sosa-Brito et al. 2001).
\ngc~also shows a sharp-edged [O III] outflow cone (Storchi-Bergmann \&
Bonatto 1991) having its vertex centered in the (supposedly) hidden nucleus of
the host galaxy.  All these pieces of evidence suggest that \ngc~has a hidden
active nucleus lying in a dusty environment and surrounded by circumnuclear
starburst.

X-ray observations of \ngc~have given further support to this scenario.  Thank
to its brightness (Piccinotti et al. 1982) this AGN has been targeted by most
of the X-ray telescopes: {\it Einstein} (Maccacaro \& Perola 1981), {\it EXOSAT}
(Turner \& Pounds 1989), {\it Ginga} (Warwick et al. 1993), {\it ASCA}
(Schachter et al. 1998, Xue et al.  1998), \beppo~(Turner et al. 2000, T00
hereafter), and \chandra~(Bianchi et al. 2007, B07 hereafter). All these
studies reported a flat X-ray spectrum dominated by heavy obscuration. The
first detailed description of the complex absorber in \ngc~was given in
T00. Exploiting the unique \beppo~broad bandpass, these authors reported the
detection of two absorbing systems: one fully-covering Compton-thin
(\nh~$\sim$ 1.4 $\times$ 10$^{23}$ \cm2) screen plus a Compton-thick
(\nh~$\sim$ 1.6 $\times$ 10$^{24}$ \cm2) absorber covering $\sim$60\% of the
X-ray source.
 Previous X-ray observations of \ngc~also revealed rapid short-term (with
a factor of $\sim$2 variations on time scale as short as 5.5 hours) as well as
complex long-term variability of the hard X-ray flux 
(e.g. Risaliti et al. 2002).

\xmm~observed this source in 2001 for 23 ks and, again, in 2005 for 100 ks.  A
partial  analysis of the first observation, limited to the hard portion (i.e.
$>$ 2 keV) of the low-resolution \epic~spectrum, can be found in Dewangan \&
Griffiths (2005).

 The analysis of the second \xmm~observation, presented in 
this paper for the first time, has provided the highest quality 
X-ray spectrum of \ngc~ever obtained. We also 
re-analyzed the 2001 dataset extending the Dewangan \&
Griffiths (2005) spectroscopic analysis to the soft X-rays range.  For both
observations we performed the analysis of the high-resolution (0.3-1 keV)
\rgs~and low-resolution (0.3-10 keV) \epic~spectroscopic data.


\section{XMM--Newton observations}

\ngc~was observed by \xmm~on 2001 May 25 and
2005 April 29 for about 23 and 100 ks, respectively.
Both observations were performed with the \epic~PN (Struder et al.
2001) and MOS (Turner et al. 2001) cameras operating in full-frame
mode.

Data were reduced with SAS 6.5 (Gabriel et al. 2003) using standard procedures.
X--ray events corresponding to patterns 0--12(0--4) for the
\mos(\pn)~cameras were selected. The event lists were filtered to
ignore periods of high  background flaring  according to the method
presented in Piconcelli et al. (2004) based on the cumulative
distribution function of background light curve count-rates. Useful
exposure times
of 71.2(18.1) ks for the \pn~and of 81(22.4) ks for the two
\mos~cameras were obtained for the 2005(2001) observation.

The source photons were extracted from a circular region with a radius of 40
arcsec (32 arcsec in the case of the 2001 observation), while the background
counts were estimated from a source-free similar region on the same
chip. Appropriate response and ancillary files for all the \epic~cameras were
created using RMFGEN and ARFGEN tasks in the SAS, respectively. As the
difference between the MOS1 and the MOS2 response matrices is a few percent,
we created a combined MOS spectrum and response matrix. The
background-subtracted spectra for the PN and the combined MOS cameras were
then simultaneously fitted.  Events outside the 0.3--10 keV range were
discarded in the PN spectrum  while  we restricted the analysis of the MOS
data to the 0.5--10 keV range\footnote{In the case of the lower statistics
data from the 2001 observation, we ignored the MOS data below 1 keV due to
the presence of cross-calibration uncertainties between the MOS cameras in the
0.3-1 keV band which could prevent an unbiased determination of the PN$+$MOS
best-fit spectral model (Stuhlinger et al. 2006)}.

The \rgs~(den Herder et al. 2001) was operating in standard spectroscopic mode
during the \xmm~observations. We have therefore reduced also the RGS data using
the standard SAS meta-task \texttt{rgsproc}, and the most updated calibration
files available at the moment the reduction was performed (October 2006). The
wavelength systematic uncertainty is 8 m\AA~across the whole RGS sensitive
bandpass.  We refer the reader to Guainazzi \& Bianchi (2007) for further
details on the data reduction of the \rgs~dataset.



\section{Temporal analysis of the 2005 observation}

X--ray light curves were extracted from the 2005 \pn~dataset in the soft
(0.2--2 keV) and hard (2--15 keV)  bands. The time bin size was set to be 5000
s for all light curves.  Background light curves were extracted from
source-free regions in the same CCD and with the same bin size. 

The soft X-ray Count Rate (CR) did not vary during the exposure apart a
small-amplitude (\simlt~20\% of the average CR) rapid drop in the bin around
$t_{exp}$ $\sim$ 85 ks.  The hard band light curve did not present any
evidence of large-amplitude variability.

The analysis of the hard X-ray variability  of the 2001 data has been recently
presented by Awaki et al. (2006).  
They reported a smooth and nearly linear
decrease (\simlt~30\%) of the hard X-ray CR from the start up to the end of
the observation, which is confirmed by our analysis.
In addition, we have also extracted the soft X-ray band
light curve for this observation.  The 0.2-2 keV CR remained almost steady
during the exposure with only a decrease of approximately 2$\sigma$ from the
average CR level in the final interval.


\section{Spectral analysis}

All spectra were analyzed with the XSPEC v11.2 package (Arnaud 1996). All
models discussed in this paper
include absorption due to the line-of-sight Galactic column of \nh~=
1.47 $\times$ 10$^{20}$ \cm2~(Stark et al. 1992).  The cosmology assumed
has $H_{\rm 0}$ = 70 km s$^{-1}$ Mpc$^{-1}$, $\Omega_{\rm m}$ =
0.3, and $\Omega_{\rm \Lambda}$ = 0.7 (Bennett et al. 2003). The quoted errors
on the model parameters correspond to a 90\% confidence level for one
interesting parameter ($\Delta\chi^2$ = 2.71; Avni 1976).

\subsection{The Hard X-ray Continuum}
\label{sec:hard}

Given the spectral complexity of \ngc, we began our analysis 
by fitting the \epic~data in the 3--10 keV range.
\begin{table*}
\caption{Spectral fitting results for the {\it dual$_{3-10}$} model applied to
the hard (3-10 keV) band. For comparison the results obtained using the 1998
\beppo~data are also listed (e.g. Sect.~\ref{sec:sax}).}
\label{tab:sax}
\begin{center}
\begin{tabular}{c c c c c c c}
\hline \hline \\
&$\Gamma$&\nh$^a$ &\nh$^b$ &$C^c_f$ & EW(Fe K$\alpha$)$^d$&$F_{3-10}$$^e$\\
& &10$^{22}$ \cm2&10$^{22}$ \cm2 & \%&eV &10$^{-12}$ \cgs\\
\hline\hline
{\it XMM-Newton}
2005&2.35$^{+0.19}_{-0.19}$&9.4$^{+1.0}_{-0.8}$&93.5$^{+6.7}_{-3.2}$&90&715$^{+365}_{-101}$&1.1\\
&&&&&\\
{\it XMM-Newton} 2001&2.28$^{+0.27}_{-0.27}$&10.2$^{+2.7}_{-3.4}$&66.9$^{+8.3}_{-4.4}$&90&330$^{+70}_{-40}$&1.7\\
&&&&&\\
\beppo~1998&2.06$^{+0.08}_{-0.14}$ &16.9$^{+1.5}_{-1.5}$&265.1$^{+115.5}_{-64.1}$&50&83$^{+23}_{-29}$ &11.7\\
\hline
\end{tabular}\end{center}
$^a$ ``fully-covering'' absorber. $^b$ ``partially-covering''
absorber. $^c$ best-fit covering factor.
 $^d$ Measured for an absorption-corrected line against absorption-corrected continuum.
 $^e$ An absorbed power-law model with $\Gamma$ = 1.7 and
\nh~= 1.5 $\times$ 10$^{22}$ \cm2 was assumed.
\end{table*}

We initially assumed the ``dual absorber'' model used by T00 for the 3--100
keV {\it BeppoSAX} data consisting of a power law modified by two neutral
absorbers, one of which (the thicker) only partially covers the primary X--ray
source.  We also included two narrow Gaussian lines in order to account for
the iron K$\alpha$ and K$\beta$ emission features, which were both
 clearly visible in the spectrum.  
The energy of the K$\beta$ line was fixed to 7.06 keV.  In the case
of the 2005 spectrum we also added in the model two narrow Gaussian lines at
6.67 and 7.55 keV to reproduce the FeXXV K$\alpha$ and NiXXI K$\alpha$
emission.  Each line represents an improvement in the resulting fit statistics
significant at 99.9\% confidence level.

This parameterization ({\it dual$_{3-10}$} hereafter) provided a very good fit
to both \xmm~datasets with a final \xndof=0.96(204) and  \xndof=0.94(172) for
the 2005 and 2001 observations, respectively (Table~1). 
The continuum slope and the column density of the
fully-covering absorber did not differ between the two epochs.  On the
contrary, the 2005 spectrum shows a much stronger \feka~line with an EW $\sim$
700 eV.  We also found a significant change in the best-fit value of the
column density of the partially--covering absorbing screen with an
increase of $\Delta$(\nhp)$\sim$ 25  $\times$ 10$^{22}$ \cm2,
while the covering factor remained unchanged ($C_f$ = 90\%)

We then adopted a model consisting of an absorbed power law plus an
unabsorbed  pure Compton reflection component from neutral matter
({\tt PEXRAV} model in XSPEC, e.g. Magdziarz \& Zdziarski 1995, with the
metal abundances fixed to the solar value, the inclination angle
fixed to 65 deg) reflecting a power-law
with the same photon index of the absorbed primary
continuum. Such a model has been found to successfully reproduce the
X--ray spectrum of most heavily--obscured Seyfert 2 galaxies
observed by \xmm~so far (e.g. Schurch, Roberts \& Warwick 2002; Matt
et al. 2004; Bianchi et al. 2005a). Narrow Gaussian lines were
included as in the  {\it dual$_{3-10}$} model. This model (indicated
as {\it refl$_{3-10}$} hereafter) gave an equally good description
of the hard X--ray spectrum of \ngc~with an associated \xndof~=
0.93(205) (\xndof~= 0.96(173) for the 2001 data). For the 2005
spectrum we measured a $\Gamma$ = 2.40$^{+0.18}_{-0.20}$ and a
column density of \nh~= (140$^{+16}_{-13}$) $\times$ 10$^{22}$ \cm2.
The energy centroid(EW) of the Fe K$\alpha$ line was 6.404$\pm$0.006
(802$^{+49}_{-45}$ eV, calculated with respect to the pure
reflection component). This energy corresponds to low ionization
states, i.e. FeI--XIII (Kallman et al. 2004).

The application of the {\it refl$_{3-10}$} model to the 2001 
hard X-ray spectrum yielded results consistent with
those reported by  Dewangan \& Griffiths (2005), i.e. $\Gamma$ =
2.08$^{+0.29}_{-0.29}$ and \nh~= (75$^{+9}_{-9}$) $\times$ 10$^{22}$
\cm2. This fit gave, therefore, a further confirmation for an
increase in the level of absorption between 2001 and 2005. On the
contrary, the values of the normalization of the reflection
component and the intensity of the \feka~line are
consistent with being the same 
in the two \xmm~observations within errors. This fact is not
surprising since both these spectral features are believed to be the
result of reprocessing of the primary (obscured) X-ray
emission in a distant material (\simgt~1 pc from the X-ray source).\\

Models {\it dual$_{3-10}$} and {\it refl$_{3-10}$} are statistically
indistinguishable. However, they correspond to very different physical scenarios
for the hard X-ray emission in \ngc, requiring either the presence of two different
absorbers (``transmission scenario'') or reflection/reprocessing from
optically thick matter (``reflection-dominated
scenario''), respectively.  We extensively discuss both scenarios and their
implications in the Sect.~\ref{sec:sax} and in Sect.~\ref{sec:scenario}.

\begin{table*}
\caption{List of the emission lines (plus OVIII RRC) detected in 
the RGS data and those included in the  \epic~best fit model
  along with their likely identification.}
\label{tab:lines}
\begin{center}
\begin{tabular}{c c c c c}
\hline \hline \\
\multicolumn{1}{c} {\bf Line Id. (\AA)} &
\multicolumn{2}{c} {\bf RGS data} &
\multicolumn{2}{c} {\bf EPIC data}\\
&$\lambda$&Intensity&$\lambda$&Intensity\\
&(\AA)&(10$^{-5}$ ph/cm$^2$/s)&(\AA)&(10$^{-5}$ ph/cm$^2$/s)\\
\hline\hline
&&&&\\
CVI Ly-$\alpha$(33.74)&33.78(33.73,33.83)&1.07(0.22,1.86)& & \\
NVII Ly-$\alpha$(24.78)&24.80(24.70,24.84)&0.68(0.28,1.11)&& \\
OVII He-$\alpha$($f$)(22.10)&22.10$^\star$&1.37(0.87,1.88)&22.10$^\dag$&1.33(1.15,1.58)\\
OVII He-$\alpha$($i$)(21.80)&21.80$^\star$&0.87(0.40,1.39)&&\\
OVII He-$\alpha$($r$)(21.60)&21.60(21.59,21.61)&0.97(0.48,1.50)&&\\
OVIII Ly-$\alpha$(18.97)&18.98(18.97,18.99)&1.47(1.13,1.85)&18.97$^\dag$&0.97(0.84,1.14)\\
OVII He-$\beta$(18.63)&18.66(18.62,18.69)&0.22(0.06,0.55)&&\\
FeXVII 3s2p(17.08)&17.10(17.08,17.11)&0.69(0.48,1.07)&&\\
FeXVIII 3s2p(16.09)&16.03(15.99,16.08)&0.45(0.21,0.67)&16.09$^\dag$&0.78(0.67,0.92)\\
OVIII Ly-$\beta$(16.00)&16.02(15.98,16.10)&0.29(0.11,0.53)&&\\
FeXVII 3d2p$^3$D$_1$(15.26)&15.26$^\dag$&0.96(0.73,1.32)&&\\
FeXVII 3d2p$^1$P$_1$(15.01)&15.01(15.00,15.03)&1.07(0.80,1.43)&14.79(14.77,14.88)&1.97(1.78,2.04)\\
FeXVIII 3d2p(14.41)&14.51(14.48,14.55)&0.32(0.09,0.60)&&\\
NeIX He-$\alpha$(r)(13.45)&13.48(13.46,13.52)&0.86(0.36,1.37)&13.45$^\dag$&1.62(1.55,1.79)\\
OVIII RRC(14.22)&14.22$^\dag$&0.30(0.11,0.71)&&\\
FeXXI 3d2p(12.28)&12.15(12.11,12.19)&0.54(0.13,0.97)&12.21(12.16,12.26)&1.30(1.15,1.35)\\
FeXXIV 3d2p(11.18)& & &11.18$^\dag$&0.41(0.34,0.47)\\
NeX Ly-$\beta$(10.24)&&&10.35(10.26,10.45)&0.29(0.21,0.33)\\
MgXI He-$\alpha$(9.23)& & &9.21(8.99,9.24)&0.46(0.38,0.49)\\
MgXII Ly-$\alpha$(8.42)& & &8.53(8.43,8.59)&0.14(0.11,0.22)\\
MgXII Ly-$\beta$(7.10)&&&7.10$^\dag$&0.09(0.06,0.13)\\
SiXIV Ly-$\alpha$(6.18)&&&6.18(6.15,6.23)&0.19(0.12,0.22)\\
SiXIII He-$\alpha$(6.69)& & &6.68(6.66,6.71)&0.48(0.43,0.56)\\
SXV He-$\alpha$(5.04)& & &5.07(5.05,5.09)&0.32(0.26,0.36)\\
SXVI Ly-$\alpha$(4.72)&&&4.68(4.60,4.76)&0.05(0.02,0.11)\\
&&&&\\
FeXXV K$\alpha$(1.86)&&&1.86$^\dag$&0.23(0.15,0.31)\\
Fe K$\beta$(1.76)&&&1.76$^\dag$&0.32(0.24,0.40)\\
NiXXI K$\alpha$(1.64)&&&1.65(1.63,1.66)&0.17(0.10,0.25)\\
\hline
\end{tabular}\end{center}
The values in parentheses in columns 2 to 5
indicate the minimum and the maximum limits
of the 90\% confidence interval for each parameter. 
$^\star$ The wavelength of the forbidden $f$ and the intercombination $i$ lines
were fixed to be at +$0.5$ and $+$0.2 \AA~from the resonance line best-fit
wavelength; $^\dag$ fixed value.
\end{table*}
\begin{figure*}
\begin{center}
\includegraphics[width=19.cm,height=9.5cm,angle=0]{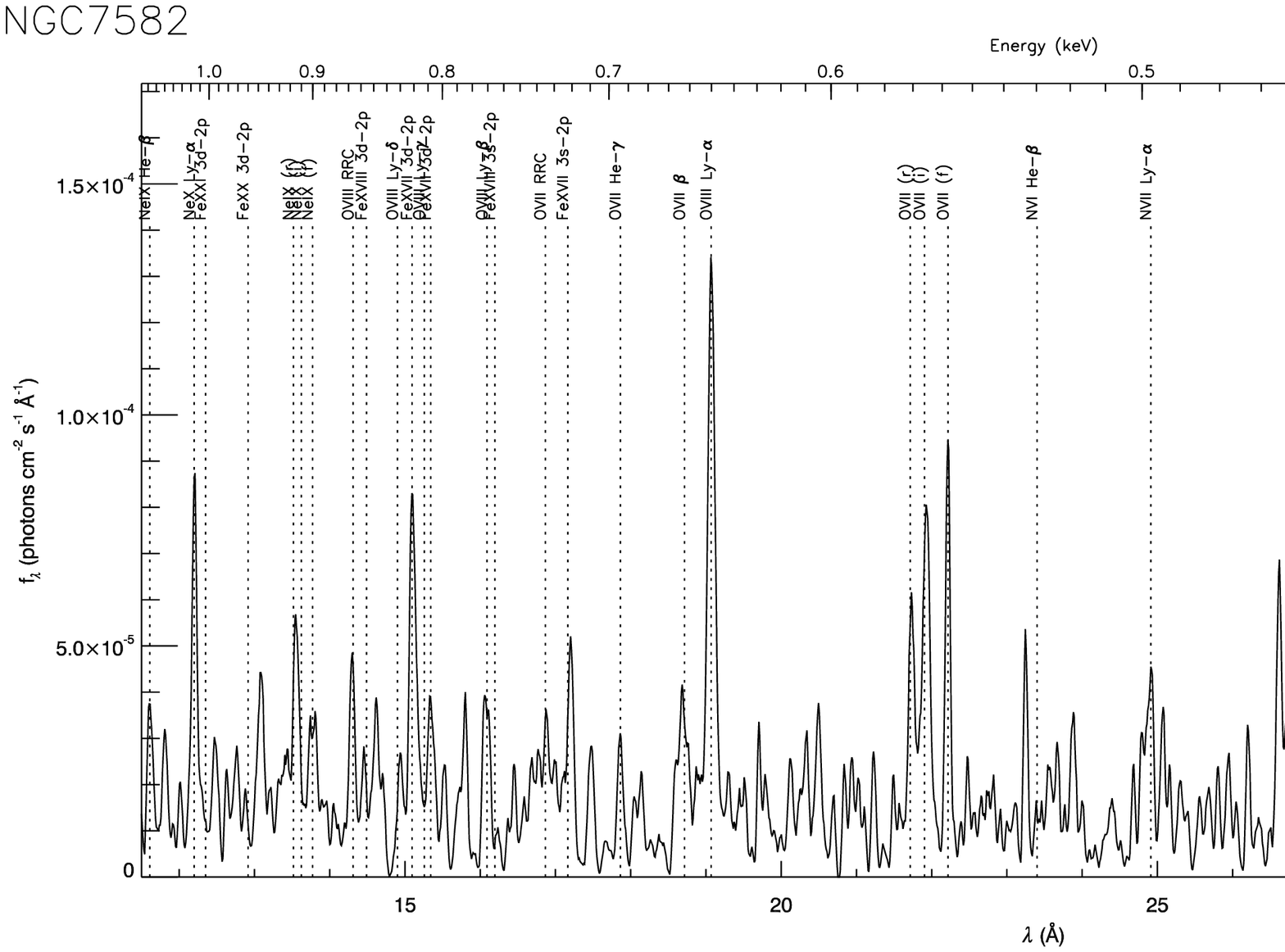}
\caption{RGS {\it fluxed} spectrum of NGC~7582 resulting from the merging of the 2005 and
  2001 \xmm~observations.  
The positions of the main emission lines produced in a photoionized plasma are labeled.
The emission lines significantly detected in the spectrum of each single RGS camera are only those
listed in Table 2 (see Sect.~\ref{sec:totale} for details).}
\label{fig:rgs}
\end{center}
\end{figure*}

\subsection{The 0.3--10 keV spectrum}
\label{sec:totale}

The extrapolation of the hard X-ray models (see Sect. \ref{sec:hard}) to
energies lower than 3 keV shows the presence of a smooth continuum excess
along with several narrow excess features.
This excess has been
observed in most of X-ray obscured Seyfert galaxies (Turner et al.
1997) and, thanks to \chandra~and \xmm~high-resolution spectroscopy,
it has been interpreted as a combination of  reflected emission
of the hidden nuclear continuum by
a warm gas  (and/or a tiny leaked
fraction of the nuclear emission) with unresolved blends of emission
lines, which originate from the same gas (e.g. Sako et al. 2000;
Sambruna et al. 2001; Matt et al. 2004; Pounds \& Vaughan 2006;
Bianchi et al. 2005a; B07).

The analysis of the RGS spectrum lends support to this interpretation,
revealing for the first time that the soft X-ray emission of \ngc~is clearly
dominated by emission lines,  as shown in Fig.~\ref{fig:rgs}.
This figure is generated by combining the 1st and
 2nd order spectra of the two RGS cameras, and smoothing the combined spectra
 through a convolution with a 5-spectral channel wide triangular function. 
Errors are not shown in the {\it fluxed} spectrum of Fig.~\ref{fig:rgs}, which is presented 
for illustration purposes only since the applied smoothing procedure can
amplify any spurious features due to instrumental issues as, for instance,
in the case of the two strong unidentified emission lines visible at 
26.5 and 28.3 \AA~that correspond 
to chip gaps in RGS2 CCD array (and are not associated with 
any relevant atomic transition).
Table 2 reports the parameters of the lines detected by a 
combined forward-folding fit of the spectra
of the two RGS cameras, employing the spectral fitting procedure
outlined in Guainazzi \& Bianchi (2007).
Their origin and the
implications concerning the physical properties of the emitting gas are
discussed in detail in Sect.~\ref{sec:soft}.

\begin{table*}
\caption{Spectral fitting results for the model \refltot~applied to the broad
  (0.3-10 keV) band.}
\label{tab:models}
\begin{center}
\begin{tabular}{c c c c c c c c c c}
\hline \hline \\
Obs.&$\Gamma$ &\nh & $A^{cont}_{pl}$&$A^{scatt}_{pl}$&$I$(Fe
K$\alpha$)& EW(Fe K$\alpha$)& Flux &Lum.&\xdof\\
(1)&(2)&(3)&(4)&(5)&(6)&(7)&(8)&(9)&(10)\\
\hline\hline
2005 &
1.93$^{+0.01}_{-0.01}$&4.8$^{+0.4~a}_{-0.6}$&35$^{+3}_{-3}$&1.06$^{+0.02}_{-0.02}$&23$^{+1}_{-1}$&772$^{+46~c}_{-36}$&3.8$^d$&41.68$^d$&381(361)\\
& &129.3$^{+5.5~b}_{-6.7}$ &&&&&23.5$^e$&41.86$^e$&\\
2001 &1.90$^{+0.05}_{-0.02}$&3.9$^{+0.5~a}_{-0.7}$&35$^{+2}_{-5}$&1.06$^{+0.04}_{-0.03}$&22$^{+2}_{-3}$&620$^{+66~c}_{-79}$&3.7$^d$&41.68$^d$&295(272)\\
& &54.9$^{+6.6~b}_{-2.1}$&&&&&40.2$^e$&41.89$^e$&\\
\hline
\end{tabular}\end{center}
The columns give the following information: (1) the
  observation date; (2) the power-law photon index; (3) the absorber column
  density of the absorber (10$^{22}$ \cm2); (4) the normalization of the
  absorbed power law (10$^{-4}$ photons/keV/cm$^2$/s); (5) the normalization
  of the scattered power law (10$^{-4}$ photons/keV/cm$^2$/s); (6)  the
  intensity of the \feka~line (10$^{-6}$ photons/cm$^2$/s); (7) the
  \feka~equivalent width (eV); (8) the flux (10$^{-13}$ \cgs); (9) the
  logarithmic value of luminosity (erg/s); (10) the reduced $\chi^2$ and the
  number of degrees of freedom.
$^a$ Absorber covering the reflected component and the primary continuum.
$^b$ Absorber covering only the primary continuum.
$^c$ Measured with respect to the pure
reflection component. $^d$ Value for the 0.5-2 keV band. $^e$ Value for the 2-10 keV band.
\end{table*}

For the above reasons,
we added a power-law component to the  {\it refl$_{3-10}$} model to
account for the continuum ``soft excess'' emission. The photon index
was constrained to be that of the primary (absorbed) continuum but
the normalization was left free to vary.  For the reasons explained in
the Sect.~\ref{sec:discussion}, model {\it refl$_{3-10}$} turns out to be
physically preferable to  model {\it dual$_{3-10}$} and we therefore used the former as
baseline model for the broad band \epic~spectrum of \ngc.
Apart from the strongest emission lines detected in the RGS spectrum in the
range 0.3--1 keV, we also included in the fitting model eight narrow Gaussian
lines, required at a significance level of $>$99\%, to account for the strongest emission features observed in the  $\sim$ 1--3 keV band, where the
RGS detectors are less sensitive.
Table~\ref{tab:lines} lists the best-fit parameters together with the
likely identification for each of these lines as measured in the longest
observation.  Finally, no absorption line is required by the data.
\begin{figure*}
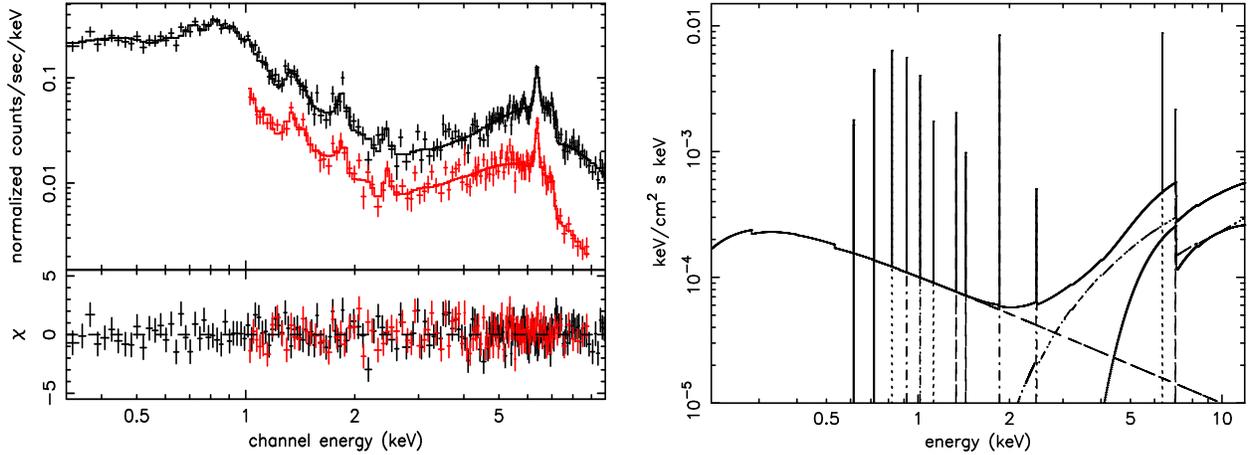

\begin{center}
\includegraphics[width=6cm,height=8cm,angle=-90]{a6439f2.ps}\hspace{0.5cm}\includegraphics[width=6cm,height=8cm,angle=-90]{a6439f3.ps}
\caption{{\it Left:} 2001 \xmm~\pn~(top) and \mos~(bottom) spectra of \ngc~when the
\refltot~model is applied.  The lower panel shows the deviations of the
observed data from the model in unit of standard deviations.  {\it Right:}
Best-fit model for the reflection-dominated scenario. This model consists of
an absorbed (\nh~$\sim$ 6 $\times$ 10$^{23}$ \cm2) power law, a pure
reflection component (both obscured by a column density of $\sim$ 4 $\times$
10$^{22}$ \cm2) plus an additional unobscured power law component accounting
for the soft X-ray scattered/leaked emission. It also includes twelve narrow
Gaussian emission lines.}
\label{fig:refltot01}
\end{center}
\end{figure*}
\begin{figure*}
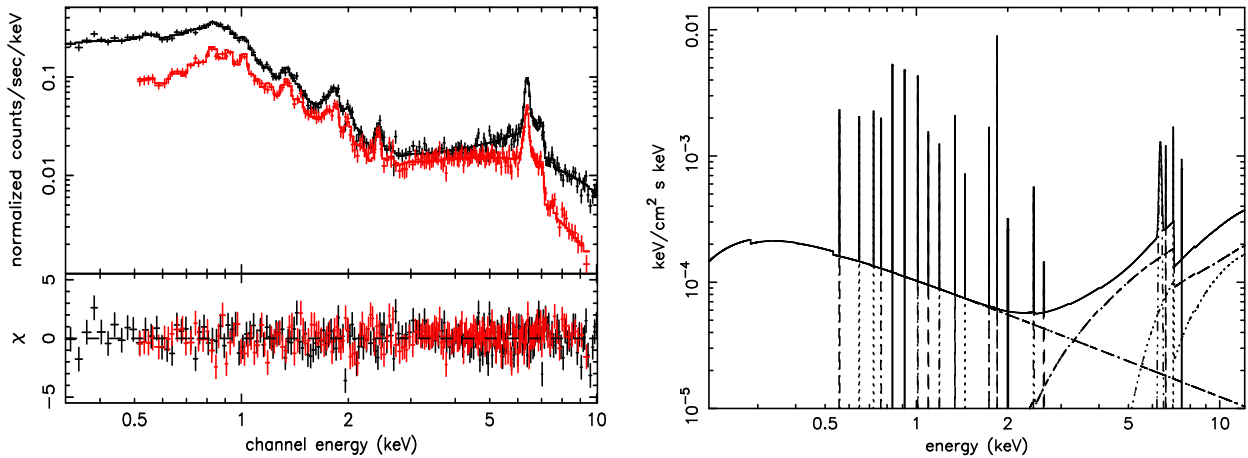

\begin{center}
\includegraphics[width=6cm,height=8cm,angle=-90]{a6439f4.ps}\hspace{0.5cm}\includegraphics[width=6cm,height=8cm,angle=-90]{a6439f5.ps}
\caption{{\it Left:} 2005 \xmm~\pn~(top) and \mos~(bottom) spectra of \ngc~when the
\refltot~model is applied.
The lower panel shows the deviations of the observed data from the model in unit of standard deviations.
{\it Right:} Best-fit model (\refltot) for the 2005 \epic~spectrum.
 See Tables~\ref{tab:lines} and \ref{tab:models} for further details.}
\label{fig:refltot05}
\end{center}
\end{figure*}

 This extension of the model {\it refl$_{3-10}$} to the broad band
produced a \xnu~\simgt~1.2 for both \xmm~datasets\footnote{Given the high
complexity of the broad band spectrum and the poorer statistics of the 20 ks
spectrum taken in 2001, we constrained the value of the normalization of the
reflection component to vary within the 99\% confidence level interval derived
for the same parameter from the 2005 data. The 2001 and 2005
unconstrained values for this parameter were
consistent within errors.}, indicating that such a parametrization of the
continuum spectral shape was not satisfactory.  However, a very significant
($\gg$99.9\% confidence level) improvement was obtained by including in this
fit  an additional absorber obscuring both the primary emission and
the Compton reflection. We derived column densities of \nh~=
(3.9$^{+0.5~}_{-0.7}$) $\times$ 10$^{22}$ \cm2~(2001 value),
and \nh~= (4.8$^{+0.4~}_{-0.6}$) $\times$ 10$^{22}$ \cm2~(2005 value). 
On the contrary, as
already pointed out in Sect.~\ref{sec:hard}, the
column density of the inner, thicker absorber
covering only the primary continuum increased by a factor
of about 2.4 ($\Delta$\nh$\approx$ 7  $\times$ 10$^{23}$ \cm2) between 2001
and 2005. The best-fit parameters yielded by the application of this model
(\refltot~hereafter) are listed in Table~\ref{tab:models}. Spectral
residuals and unfolded spectral models are shown in
Figs.~\ref{fig:refltot01} and ~\ref{fig:refltot05}.
\begin{figure*}
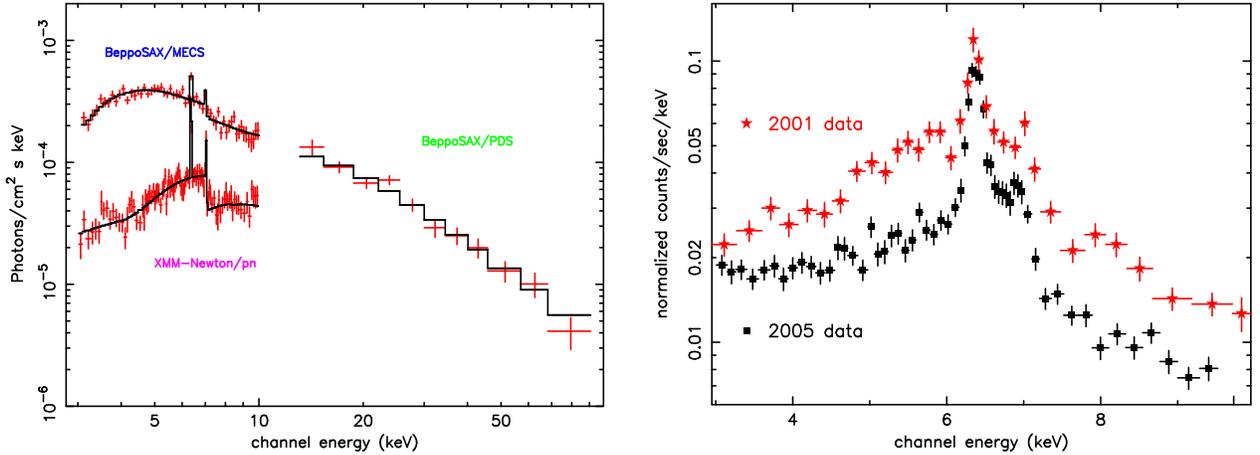

\begin{center}
\includegraphics[width=6cm,height=8cm,angle=-90]{a6439f6.ps}\hspace{0.5cm}
\includegraphics[width=6cm,height=8cm,angle=-90]{a6439f7.ps}
\caption{(a) {\it Left:} Unfolded  1998 \beppo~(MECS and PDS) and 2001 \xmm~\pn~spectra of \ngc~fitted
  by the dual$_{3-10}$ model (e.g. Sect.~\ref{sec:sax} for details). This
  figure clearly shows the large variations in the spectral shape and
  intensity occurred in the source between the two observations.  (b) {\it
  Right:} 3--10 keV PN data from the observations of 2001 May and 2005
  April. This plot demonstrates the progressive decline of the hard X-ray flux.}
\label{fig:sax}
\end{center}
\end{figure*}


\section{\xmm~versus \beppo~results: evidence of a spectral transition}
\label{sec:sax}

In Fig.~\ref{fig:sax}a the 2001 3--10 keV \pn~spectrum and the 3--90
keV {\it MECS}$+${\it PDS} spectra are plotted together and fitted by
the best-fit model used by T00 for the continuum emission above 3
keV (which is similar to our {\it dual$_{3-10}$} model).
It is worth noting the decrease of the X-ray flux between
the two epochs:
 \xmm~caught, in fact, the source in a extremely
faint state with a 3--10(8--10) keV flux lower than a factor of
$\sim$7(4) if compared with the \beppo~observation.
\ngc~was even fainter during the second \xmm~observation performed
four years later as shown in Fig.~\ref{fig:sax}b.
Among the previous studies of \ngc, the \xmm~continuum flux levels
(e.g. $F^{2001}_{2-10}$ = 4 $\times$
10$^{-12}$ \cgs~and $F^{2005}_{2-10}$ = 2.3 $\times$ 10$^{-12}$ \cgs) are the lowest
ever measured: {\it EXOSAT} (\fhx~$\sim$ 1.7 $\times$ 10$^{-11}$
\cgs, Turner \& Pounds 1989),
{\it ASCA} (\fhx~$\sim$ 1.5 $\times$ 10$^{-11}$ \cgs, Xue et
al. 1998) and {\it Ginga}  (\fhx~$\sim$ 6 $\times$ 10$^{-12}$ \cgs,
Warwick et al. 1993) all observed the source in a brighter state.

Spectral variations are also evident between the \beppo~and
\xmm~observations (see Table~\ref{tab:sax} and Fig.~\ref{fig:sax}).
The \xmm~spectrum of \ngc~shows the
typical features (i.e. a strong iron line emerging from a hard and
curved continuum) of a source in a ``reflection-dominated" phase.
The detection of a prominent \feka~line (EW$\sim$700 eV; see
Table~\ref{tab:sax}) coupled with the excellent fit to the data provided by
the model {\it refl$_{3-10}$} (e.g. Sect.~\ref{sec:hard}) suggest
that the X-ray emission detected by \xmm~can be considered as the
echo of primary -- {\it i.e.} nuclear -- continuum during a prior
phase of more intense nuclear activity. This change of appearance
from a transmission- to reflection-dominated spectral state (or
vice versa, e.g. Guainazzi et al. 2002) was reported by Matt,
Guainazzi \& Maiolino (2003) and Bianchi et al. (2005b) in a handful
of Seyfert 2 galaxies. They argued that this temporal behavior can
be due to switching-off (or following re-emergence) of the nuclear activity.

\section{Discussion}
\label{sec:discussion}

\subsection{A possible geometry for the nuclear region in \ngc}
\label{sec:scenario}

In this section we  focus on the properties of the very complex X-ray spectrum
above $\sim$3 keV with the aim of providing a physically acceptable scenario
for the nuclear emission of \ngc.

Both the {\it refl$_{3-10}$} and the {\it dual$_{3-10}$} model describe the
data equally well (see Sect.~\ref{sec:hard}). However, the comparison between
the \beppo~and {\it XMM-Newton} X-ray spectra leads us to favor the former on
the basis of physical considerations.  In a
pure transmission-dominated scenario with the 2005(2001) X-ray continuum
attenuated  by a column of \nh~$\sim$ 9(7) $\times$ 10$^{23}$ \cm2~there is 
no explanation for the presence of a strong \feka~line with an
EW of $\sim$~700 (330) eV (e.g. Matt 2002).  It is also  worth noting that
during the \xmm~exposures of \ngc~no rapid and large-scale variations in the
hard band light curve  occurred, at odds with previous
\asca~and \beppo~observations which caught the source in a clear
transmission-dominated phase.  This suggests that the primary X-ray
source has experienced a large-amplitude decline prior to the 2001
\xmm~observation unmasking the reflection component
underlying the primary emission.
Hereafter we therefore refer to the reflection-dominated spectral state model
(i.e. \refltot) as the most physically acceptable description of the
\xmm~broad band spectra of \ngc.
\begin{figure}
\begin{center}
\includegraphics[width=8cm,height=8cm,angle=0]{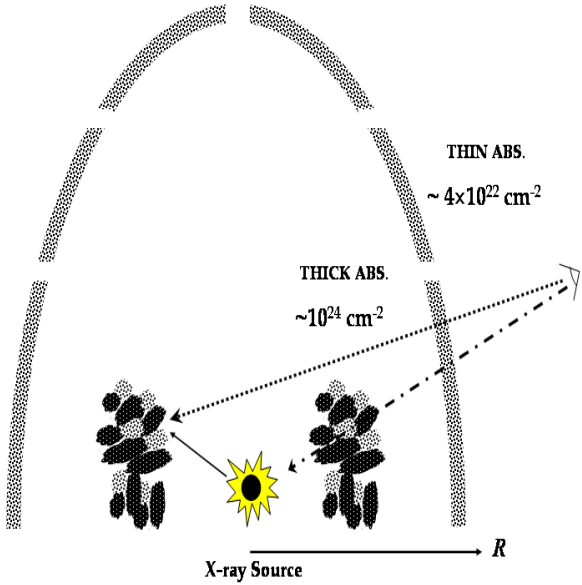}
\caption{Sketch of the absorption geometry in the circumnuclear region of
  \ngc~according to model \refltot. The {\it dotted} and {\it dotted-dashed}
  lines represent the X-ray lines of sight.  A fraction of the X-ray continuum
  intercepts ({\it solid line})  the inner surface of the thick (\nh~$\sim$
  10$^{24}$ \cm2) absorber and is reflected via Compton scattering along our
  line of sight.  These photons scattered in our direction will then pass
  through the large-scale thin absorber, which is located far away from the
  central X-ray source, and be absorbed by a column density of \nh~$\sim$ 4
  $\times$ 10$^{22}$ \cm2. This figure is
  not to scale.}
\label{fig:sketch}
\end{center}
\end{figure}

A similar apparent switching-off of the nucleus was observed in other
Seyfert 2 galaxies (e.g. Guainazzi et al. 2005, G05 hereafter) and even in
Type 1 AGN (Guainazzi et al. 1998).
The physical reason of this phenomenon is
likely associated with variations in the accretion flow onto the BH
(Uttley et al. 1999)
or,  at
least in some cases, with dramatic changes in the properties
(i.e. column density, covering factor) of the line-of-sight absorber
(e.g. Risaliti et al. 2005; Elvis et al. 2004; Matt
Guainazzi \& Maiolino 2003). It is worthwhile to notice that
in \ngc~the AGN
continuum has not completely disappeared as a significant component
of transmitted (absorbed) nuclear emission is  detected (e.g.
Figs.~\ref{fig:refltot01}, \ref{fig:refltot05}  and Table~\ref{tab:models}).

Our best-fit model for both \xmm~observations requires the presence
of two absorbers (e.g. Table~\ref{tab:models}): a screen with
\nh~$\approx$ 4 $\times$ \due~\cm2, which blocks both primary and
reflected emission, and a medium with \nh~$\approx$ 5--10 $\times$
\tre~\cm2 which obscures the nuclear source only. In  the following
we will refer to them as the thin and the thick absorber,
respectively.
The gas responsible for the reflection is most likely located at the
far inner side of the molecular {\it torus} (i.e. the
cylindrically-symmetric compact dusty absorber invoked in the
simplest version of the AGN unified models, e.g. Krolik \& Begelman
1988; Antonucci 1993; see also Jaffe et al. 2004).  Given the column
density of the thick absorber, it is quite straightforward to
associate it to the {\it torus}. This in turn
implies that the thin absorber surrounds both the central
engine and reflecting/reprocessing region and, therefore, should be
located well beyond the pc-scale distance usually attributed to the
{\it torus}. A sketch of the proposed absorption geometry in
the circumnuclear region of NGC 7582 is shown in
Fig.~\ref{fig:sketch}.

This scenario is also supported by the discovery of
an increase of a factor of $\sim$2.4 in the column density 
of the thick absorber  between the two \xmm~observations
(i.e. over an interval of about 4 years). Interestingly, similar
large-amplitude variations of the absorbing column over the whole X-ray
history of \ngc~were reported by Risaliti et al. (2002).
They measured a fast variation in the \nh~value by a
factor of 1.2 in a time interval of six months. The \nh~change
observed between the \xmm~observations is, however, one of the
largest amplitude detected so far. As suggested by Risaliti et al. (2002)
and Matt, Guainazzi \& Maiolino (2003), these variations cast
serious doubts on the existence of a single homogeneous absorber
(i.e. {\it torus}) as invoked in the simplest version
of the unified models for AGN. They instead support a more complex
scenario whereby the obscuring gas is largely inhomogeneous in space or
time, with multiple absorbing components distributed in a range of
distances from fractions of parsec up to several hundreds
parsecs from the central
supermassive BH.

Our results also corroborate the conclusions drawn by B07 on the basis of
\chandra~and {\it HST} images of \ngc. The extremely high value of the nuclear
flux ratio between 1.6 and 0.6 $\mu$m measured for this Seyfert galaxy
suggests, in fact,  the presence of a second intervening absorber more external
than the expected {\it torus}. This could be linked to a
prominent dust lane and/or dusty compact regions associated with starburst
formation (e.g. Sosa-Brito et al. 2001; Wold \& Galliano 2006) in this source.
The \xmm~observations presented in this paper provide us with the direct evidence of
the existence of two circumnuclear absorbing regions and allow us to shed light
on their location.

This finding also agrees with  the results presented by Guainazzi, Matt \&
Perola (2005) on the basis of a large sample of Seyfert 2 galaxies observed
with \chandra~or \xmm~and {\it HST} high-resolution optical coverage.  They
found that the presence of dust lane is correlated with values of the X-ray
column density in the Compton-thin range.  Even if the large-scale dust lanes
seen in  {\it HST} images do not seem to be directly responsible for the X-ray
obscuration, these authors suggested that a dusty inner host galaxy
environment may favor the formation of clumps/filaments of X-ray
Compton-thin absorbing matter with large covering fractions.

These results lend support to the model presented by Matt (2000).  He proposed
a modification of the simplest version of the unification model for AGN
whereby the line of sight of Compton-thick Seyfert 2 galaxies intercepts a
compact distribution of dense matter (the {\it torus}) close (\simlt~1--10 pc)
to the nucleus, while Compton-thin Seyfert galaxies are viewed through
absorbing patches likely associated with the dust lane(s) extending on scales
of hundreds of parsecs observed in all Seyfert galaxies (see Malkan et
al. 1988; Maiolino \& Rieke 1995).  As predicted by this model, in the case of
\ngc~our line of sight intercepts both the {\it torus} and the
``Compton-thin'' regions which, therefore, coexist in this source.

\subsubsection{The Fe K$\alpha$ emission line}

The EW values of the \feka~line at 6.4 keV derived from the \xmm~observations,
i.e. EW$^{2001}$ $\sim$ 330 and EW$^{2005}$ $\sim$ 700 eV 
assuming a transmission-dominated scenario (see Table 1),
are by far the largest measured
in \ngc~so far. All previous X-ray observations reported a  weaker
feature with an EW \simlt~150 eV.
 However, when calculated with respect to the reflection
component, the 2001 value of the line
EW of $\sim$600 eV (e.g. Table 3) appears to be relatively
low if compared with the expected one, i.e $\sim$1 keV (Matt et al. 1996),
 and usually
observed in reflection-dominated sources (G05). Nonetheless, an iron
K line with a EW$\sim$600 eV has been recently detected in the
mildly Compton-thick Seyfert galaxy Mrk 3 by Bianchi et al. (2005a).
They suggested to explain such an \feka~line intensity by the
combination of iron under-abundance and a small inclination angle
(i.e. $\ll$ 90 deg) between our line of sight and the symmetry axis
of reflecting medium. This interpretation might hold also for \ngc.

However, it is also worth noting that nearly all the
reflection-dominated sources observed to date are truly
Compton-thick, i.e. obscured by a \nh~\simgt~$\sigma^{-1}_t$
$\approx$ 1.6 $\times$ 10$^{24}$ \cm2, while the 2001 \xmm~spectrum
of \ngc~turns out to be absorbed by a smaller column density (i.e.
\nh~$\sim$ 5.5 $\times$ 10$^{23}$ \cm2). Furthermore, in most of the
theoretical calculations the geometry of the reflector is assumed to
be a parallel slab illuminated by  an isotropic X-ray source and
this assumption may be incorrect in the case of \ngc. In fact, as
emerged from this paper, the properties of the nuclear region in
this Seyfert galaxy seems to be peculiar, with a lot of (variable)
components contributing to the complex observed spectrum.

\subsection{The soft X-ray emission}
\label{sec:soft} The combination of \epic$+$\rgs~data provides by
far the best quality spectrum of the soft X-ray emission in \ngc~to
date.  Our spectral analysis of these data has revealed that this
emission is dominated  by a wealth of emission lines from different
ions (Fig.~\ref{fig:rgs}) superimposed on a very faint (i.e.
\simlt~2\% of the primary continuum) scattered continuum.

Table~\ref{tab:lines} shows the most prominent emission lines
detected both in the \rgs~and in the \epic~spectra. All the line
widths are unresolved and most of them were identified with
hydrogen- and helium-like lines of the most abundant metals, from
 carbon to sulfur. The strongest lines are associated to 
the OVII He-$\alpha$ and OVIII Ly-$\alpha$ transitions. An intense
OVIII Ly-$\alpha$ is usually found in galaxies with powerful
starburst activity (e.g. Guainazzi \& Bianchi 2007). However, B07
found no spatial correlation between the peaks of the OVIII emission
and the optical star-forming regions in the high-resolution composite
\chandra~and {\it HST}~image of \ngc. Furthermore, the  detection of
the narrow OVIII Radiative Recombination Continuum (RRC) can be
interpreted as strong evidence for recombination in a
low-temperature photoionized plasma. In fact, in a collisional
plasma these features are broad and weak and, therefore, hardly
discernible above the bremsstrahlung continuum (Liedahl 1999). The
presence of Fe L (i.e.  FeXVII-XVIII) features can also provide
further information on the nature of the line-emitting gas.
 The larger intensity of the 3$d$-2$p$ lines with respect to the
3$s$-2$p$ ones is at odds with the expectations for a
recombination-dominated plasma, suggesting 
an important contribution from 
photoexcitation of bound-bound transitions by the
continuum radiation field (Kallman et al. 1996; Sako et al.
2000).

This discovery significantly improves our understanding of the nature of the
``soft excess'' emission in this Seyfert galaxy.  In fact, previous \asca~and
\beppo~observations did not provide firm conclusions on the origin of this
spectral component.  T00 invoked the presence of emission from a collisionally
ionized gas ({\tt mekal} model in XSPEC), accounting for 41\% of the total
0.1-2 keV flux, plus a ``leaking'' unabsorbed fraction ($\sim$ 0.4\%) of the
hard X-ray continuum power law. Xue et al. (1998) suggested a mixture of
scattered and thermal emission without detecting any emission line due to the
poor statistics of their \asca~data.

A detailed discussion on the soft X-ray emission morphology of \ngc~has
been presented in B07 on the basis of high-resolution \chandra~and {\it HST}
images. According to B07, the hard X-ray \chandra~image is dominated by the
nuclear unresolved emission while the soft X-ray emission is extended over
$\sim$ 20 arcsec.  The large-scale dust lane visible in the optical image
strongly affects the morphology of this soft X-ray  emission. In particular,
most of it arises on the west side of the nucleus where the dust lane 
is optically
thinner and, remarkably, exhibits similarity with the [O III] radiation
cone-shaped structure detected by Storchi-Bergmann \& Bonatto (1991).

BO6 also present evidence for  two ``hot spots'' i.e. regions where
emission from higher ionization stages of O and Ne is enhanced.
These ``hot-spots'' appear to be unrelated with
nuclear star-forming regions, and therefore suggest that the
star-forming activity does not provide an important contribution to
the ionization of the soft X-ray emitting gas. Such a conclusion can
be also inferred by our analysis. The far infrared (FIR)
luminosity of \ngc~is $L_{FIR}$ $\sim$ 9.6 $\times$ 10$^{43}$
\ergs~(Taniguchi \& Ohyama 1998). Using the relationship
between FIR and soft X-ray luminosity discovered by Ranalli et al.
(2003) on a large sample of star-forming galaxies, we estimate a
starburst contribution \simlt~5\% to the total soft X-ray luminosity
in \ngc~(i.e. \lums~= 4.8 $\times$ 10$^{41}$ \ergs).

\section{Summary}
We have presented the spectral analysis of the
\epic$+$\rgs~data from two  \xmm~observations of the bright Seyfert
2 galaxy \ngc~(May 2001 and April 2005).
The main results from our analysis can be summarized as follows:\\

$-$~The spectrum of the broadband continuum emission is very
complex. It can be well described by  a  model consisting of  a
combination of a heavily absorbed (\nh~$\sim$ 12.9(5.5) $\times$
10$^{23}$ \cm2~in the 2005(2001) spectrum) power law and a pure
reflection component, both obscured by an additional
absorber with column density of $\sim$ 4
$\times$ 10$^{22}$ \cm2.

We consider this model as the most physically plausible description for the
 \xmm~data on the basis of the following observational pieces of evidence.
 Firstly, the hard X-rays light curves do not show any  rapid and/or large
 amplitude variation which are typically observed in
 ``transmission-dominated'' Seyfert 2 galaxies and, in particular, in all
 previous observations of \ngc.  Noteworthy, the 2--10 keV flux level during
 both \xmm~observations (i.e. {$F_{2-10}^{2005}$}({$F_{2-10}^{2001}$}) $\sim$
 2(4) $\times$ 10$^{-12}$ \cgs) are by far the lowest values measured for this
 source to date.  Secondly, comparing the \xmm~spectra with that obtained from
 a 1998 \beppo~observation (see Fig.~\ref{fig:sax}),  it appears evident that
 \ngc~experienced a dramatic spectral transition. 
The most prominent spectral feature
 above 3 keV is the strong narrow \feka~line at 6.4 keV (EW$^{2005}$ $\sim$
 700 eV).   
All these pieces of evidence lend support to a scenario according to
 which \xmm~caught the source in a Compton reflection-dominated phase with a
 very faint level of the (absorbed) primary continuum and a strong reflection
 component.\\

$-$~Our best-fit model requires the presence of two absorbers, with
the thinner (i.e. \nh~$\sim$ 4 $\times$ 10$^{22}$ \cm2) 
one obscuring both primary and reflected emission.  This
finding further strengthens the conclusions reported by Guainazzi,
Matt \& Perola (2005) and Matt,  Guainazzi \& Maiolino (2003) about
a distribution of the X-ray absorbing gas much more complex   than
that usually assumed in the most popular Unified models for AGN. In
particular, the X-ray spectrum of \ngc~confirms the presence of a
multiplicity of  absorbing regions coexisting in the same source.
We have detected an unprecedented increase in the column density of the
inner, thicker absorber of $\Delta$\nh~$\sim$ 7.4  $\times$ 10$^{23}$
\cm2~between 2001 and 2005, while, on the contrary,  the \nh~of the second
absorber remained unchanged.  This huge  variation suggests
strong clumpiness and inhomogeneity of the absorbing matter along the line of
sight.  A likely explanation for the observed spectral variability is the
drift of clouds  located very near the central source as proposed by
Risaliti et al. (2005) and Lamer et al. (2003) to account for similar changes
in the obscuring column density of  NGC 1365 and NGC 3227, respectively.

 Future observations above 10 keV of
\ngc~will be crucial to shed light on the properties of
the X-ray continuum emission as well as the geometry and variability
pattern of the obscuring gas in the nuclear
environment of this peculiar Seyfert 2 galaxy.\\

$-$~ The analysis of the high resolution RGS spectrum has revealed
that soft X-ray excess in this obscured AGN is dominated by a wealth
of emission lines with a very low level of scattered/leaked
continuum. According to the recent \chandra~results presented by
B07, this emission is extended on a hundreds-pc scale and,
remarkably, its shape and location are very similar to the optical
[OIII] ionization cone observed by Storchi-Bergmann \& Bonatto
(1991). The most prominent emission lines originate from H- and
He-like ions of O, C and Ne as well as by L-shell transitions of
FeXVII. The detection of a narrow OVIII RRC strongly suggests that
most of the soft X-ray emission arises in a photoionized plasma.
These findings are in agreement with the few high-resolution soft
X-ray spectroscopic measurements published to date (e.g. Sambruna et
al. 2001; Young et al. 2001; Kinkhabwala et al. 2002; Guainazzi \&
Bianchi 2007).

\begin{acknowledgements}

We would like to thank the staff of the \xmm~Science Operations Center for
their support. Useful discussions with Fabrizio Nicastro are acknowledged.
This research has made use of the NASA$/$IPAC
Extragalactic Database (NED) which is operated by the Jet Propulsion
Laboratory,  California Institute of Technology, under contract with
the National Aeronautics and Space Administration.
EP acknowledges the financial support from INAF.
\end{acknowledgements}


\begin{thebibliography}{}
\bibitem {} Antonucci, R., 1993, ARA\&A, 31, 473
\bibitem {} Arnaud  K.A., 1996, ASP Conf. Series, 101, 17
\bibitem {} Awaki, H., Murakami, H., Ogawa, Y., Leighly, K. M., 2006, ApJ, 645,
  928
\bibitem {} Bennett, C.L., et al., 2003, ApJS, 148, 1
\bibitem {} Bianchi, S., Miniutti, G., Fabian, A. C., Iwasawa, K., 2005a, MNRAS, 360, 380
\bibitem {} Bianchi, S., Guainazzi, M., Matt, G., et al., 2005b, A\&A, 442, 185
\bibitem {} Bianchi, S., Guainazzi, M., Chiaberge, M., 2006, A\&A, 448, 499
\bibitem {} Bianchi, S., Chiaberge, M., Piconcelli, E., Guainazzi, M., 2007, 374, 697 (B07)
\bibitem {} Cid Fernandes, R.., Storchi-Bergmann, T., Schmitt, H. R., 1998,
  MNRAS, 297, 579
\bibitem {} den Herder, J. W., et al.,  2001, A\&A, 365, L7
\bibitem {} Dewangan, G. C., Griffiths, R. E., 2005, ApJ, 625, L31
\bibitem {} Elvis, M., Risaliti, G., Nicastro, F., et al., 2004, ApJ, 615, L25
\bibitem {} Gabriel C., Denby M., Fyfe D. J., Hoar J., Ibarra A., 2003, in ASP Conf. Ser., Vol. 314 Astronomical Data Analysis Software and Systems XIII, eds. F. Ochsenbein, M. Allen, \& D. Egret (San Francisco: ASP), 759 
\bibitem {} Gu, M. F., Kahn, S. M., Savin, D. W., et al., 1999, ApJ, 518, 1002
\bibitem {} Guainazzi, M., et al., 1998, MNRAS, 301, L1
\bibitem {} Guainazzi, M., Matt, G., Fiore, F., Perola, G. C., 2002, A\&A, 388, 787
\bibitem {} Guainazzi, M., Matt, G., Perola, G. C.,  2005, A\&A, 444, 119
\bibitem {} Guainazzi, M., Fabian, A.C., Iwasawa, K., et al., 2005, MNRAS, 356, 295 (G05)
\bibitem {} Guainazzi, M., \& Bianchi, S., 2007, MNRAS, 374, 1290
\bibitem {} Jaffe, W., et al., 2004, Nature, 429, 47
\bibitem {} Jimenez-Bailon, E., et al., 2003, ApJ, 593, 127
\bibitem {} Kallman, T. R., Liedahl, D., Osterheld, A., Goldstein, W., Kahn,
  S., 1996, ApJ, 465, 994
\bibitem {} Kallman, T. R., Palmeri, P., Bautista, M. A., Mendoza, C., Krolik,
  J. H., 2004, ApJS, 155, 675
\bibitem {} Kinkhabwala, A., Sako, M., Behar, E., et al.,  2002, ApJ, 575, 732
\bibitem {} Krolik, J. H., \& Begelman, M. C.,  1988, ApJ, 329, 702
\bibitem {} Lamer, G., Uttley, P., McHardy, I. M., 2003, MNRAS, 342, L41
\bibitem {} Liedahl, D. A. 1999, in X-Ray Spectrocopy in Astrophysics, ed. J. van Paradijs \& J. A. M. Bleeker (Berlin: Springer), 189
\bibitem {} Maccacaro, T., Perola, G. C., 1981, ApJ, 246, L11
\bibitem {} Magdziarz, P., \& Zdziarski, A.,  1995, MNRAS, 273, 837
\bibitem {} Maiolino, R., \& Rieke, G. H.,  1995, ApJ, 454, 95
\bibitem {} Makishima, K., 1986, Lecture Notes in Physics, 266, 246
\bibitem {} Malkan, M. A., Gorjian, V., Tam, T., 1998, ApJS, 117, 25
\bibitem {} Matt, G.,  Brandt, W. N., Fabian, A. C., 1996, MNRAS, 280, 823
\bibitem {} Matt, G., 2000, A\&A, 355, L31
\bibitem {} Matt, G., 2002, MNRAS, 337, 147
\bibitem {} Matt, G., Guainazzi, M., Maiolino, R.,  2003, MNRAS, 342, 422
\bibitem {} Matt, G., Bianchi, S., D'Ammando, F., Martocchia, A., 2004, A\&A, 421, 473
\bibitem {} Molendi, S., Bianchi, S., Matt, G.,  2003, MNRAS, 343, L1
\bibitem {} Piccinotti, G., Mushotzky, R. F., Boldt, E. A., et al., 1982, ApJ, 253, 485
\bibitem {} Piconcelli, E., Jimenez-Bailon, E., Guainazzi, M., et al., 2004, MNRAS, 351, 161
\bibitem {} Pounds, K., \& Vaughan, S., 2006, MNRAS, 368, 707
\bibitem {} Prieto, M. A., Reunanen, J., Kotilainen, J. K., 2002, ApJ, 571, L7
\bibitem {} Ranalli, P., Comastri, A., Setti, G., 2003, A\&A, 399, 39
\bibitem {} Regan, M. W., \& Mulchaey, J. S,  1999, AJ, 117, 2676
\bibitem {} Risaliti, G., Elvis, M., Nicastro, F., 2002, ApJ, 571, 234
\bibitem {} Risaliti, G., Elvis, M., Fabbiano, G., Baldi, A., Zezas, A., 2005
  ApJ, 623, L93
\bibitem {} Sako, M., Kahn, S. M., Paerels, F., Liedahl, D. A., 2000, ApJ, 543, L115
\bibitem {} Sambruna, R. M., Netzer, H., Kaspi, S., et al., 2001, ApJ, 546, L13
\bibitem {} Schachter, J. F., Fiore, F., Elvis, M., et al., 1998, ApJ, 503, L123
\bibitem {} Schurch, N. J., Roberts, T. P., Warwick, R. S., 2002, MNRAS, 335, 241
\bibitem {} Sosa-Brito, R, M., Tacconi-Garman, L. E., Lehnert, M. D.,
  Gallimore, J., 2001, ApJS, 136, 61
\bibitem {} Stark, A. A., et al., 1992, ApJS, 79, 77
\bibitem {} Storchi-Bergmann, T., \& Bonatto, C. J., 1991, MNRAS, 250, 138
\bibitem {} Struder, L., Briel, U., Dennerl, K., et al., 2001, A\&A 365, L18
\bibitem {} Stuhlinger M., et al., 2006, XMM-SOC-CAL-TN-0052, Issue 3.0,
  (astro-ph/0511395)
\bibitem {} Taniguchi, Y., \& Ohyama, Y., 1998, ApJ, 507, L121
\bibitem {} Turner, T. J., \&  Pounds, K. A., 1989, MNRAS, 240, 833
\bibitem {} Turner, T. J., George, I. M., Nandra, K., Mushotzky, R. F., 1997,
  ApJS, 113, 23
\bibitem {} Turner, T. J., Perola, G. C., Fiore, F., et al., 2000, ApJ, 531, 245 (T00)
\bibitem {} Turner, M.~J.~L., Abbey, A., Arnaud, M., et al., 2001, A\&A, 365,
  L27
\bibitem {} Uttley, P., McHardy, I. M., Papdakis, I. E., Guainazzi, M.,
  Fruscione, A., 1999, MNRAS L6
\bibitem {} Ward, M. J., Wilson, A. S., Penston, M. V., et al.,   1978, ApJ,
  223, 788
\bibitem {} Warwick, R. S., Sembay, S., Yaqoob, T.,  etal., 1993, MNRAS, 265, 412
\bibitem {} Wold, M., \& Galliano, E., 2006, MNRAS, 369, L47
\bibitem {} Xue, S., Otani, C., Mihara, T., Cappi, M., Matsuoka, M.,  1998,
  PASJ, 50, 519
\bibitem {} Young, A. J., Wilson, A. S., Shopbell, P. L., 2001, ApJ, 556, 6
\end{thebibliography}
\end{document}